\documentclass[conference]{IEEEtran}
\usepackage{cite}
\usepackage{amsfonts}

\usepackage{graphicx}
\ifCLASSINFOpdf

\else

\fi

\begin{document}
\IEEEoverridecommandlockouts
\title{Energy Efficient Iterative Waterfilling for the MIMO Broadcasting Channels\thanks{This work is supported in part by Huawei
Technologies, Co. Ltd., Shanghai, China and National Basic Research Program of China (973 Program) 2007CB310602.}\thanks{Corresponding author:
Ling Qiu, lqiu@ustc.edu.cn.}}

\author{\IEEEauthorblockN{Jie Xu, Ling Qiu}
\IEEEauthorblockA{
University of Science and Technology of China (USTC),\\
Hefei, China, 230027\\
Email: suming@mail.ustc.edu.cn, lqiu@ustc.edu.cn} \and \IEEEauthorblockN{Shunqing Zhang} \IEEEauthorblockA{Huawei
Technologies, Co. Ltd.,\\ Shanghai, China\\
Email: sqzhang@huawei.com}}


%


\maketitle

\begin{abstract}
Optimizing energy efficiency (EE) for the MIMO broadcasting channels (BC) is considered in this paper, where a practical power model is taken
into account. Although the EE of the MIMO BC is non-concave, we reformulate it as a quasiconcave function based on the uplink-downlink duality.
After that, an energy efficient iterative waterfilling scheme is proposed based on the block-coordinate ascent algorithm to obtain the optimal
transmission policy efficiently, and the solution is proved to be convergent. Through simulations, we validate the efficiency of the proposed
scheme and discuss the system parameters' effect on the EE.
\end{abstract}


%
\IEEEpeerreviewmaketitle

\section{Introduction}

Wireless communication turns to the era of green. This is not only because of the exponential traffic growth with the popularity of the smart
phone but also the limited energy source with ever higher prices. Energy efficiency (EE), as a result, becomes one of the major topics in the
research of wireless communications \cite{YChenComMag} and plenty of research projects either government funded or industrial funded start to
investigate the energy efficient solutions for the wireless network as well as the sustainable future for the wireless communications.
Meanwhile, multiple input multiple output (MIMO), especially downlink multiuser MIMO (also called MIMO broadcasting channels, BC), has become a
key technology in the cellular networks due to its significant spectral efficiency (SE) improvement. Therefore, studying the EE
of the MIMO BC is a critical issue.


The EE is in general defined as the capacity divided by the power consumption, which denotes the delivered bits per-unit energy measured in bits
per-Joule. There are a lot of literatures discussing the EE of the point to point MIMO channels
\cite{Cui,chongCJ11,sprabhuenergyefficient,Kim2,EEPrecoding,AccurateEEMIMO}. The point to point MIMO channels can always be separated into
parallel sub-channels through singular value decomposition (SVD) or after detection. In this case, only power allocation across the sub-channels
needs to be optimized to compromise the transmit power and circuit power, and thus maximize the EE
\cite{chongCJ11,sprabhuenergyefficient,Kim2}. As the sub-channels are parallel, the solution is similar with the energy efficient power
allocation in OFDM systems \cite{Miao3,sprabhuenergyefficient2}. The optimization for point to point MIMO channels is not applicable for the
MIMO BC, as the MIMO BC cannot be simply transformed into parallel sub-channels{\footnote{Although after zero-forcing (ZF) precoding, for
example, the MIMO can be separated into parallel sub-channels, the ZF scheme is far away from the optimal solution \cite{DPC}.}}. There are few
literatures discussing the EE for the MIMO BC. To the best of the authors' knowledge, only \cite{chong2011VTC} and our previous work \cite{Xu1}
addressed this topic, but they both assumed linear precoding design and equal transmit power allocation for simplification. The assumption of linear precoding makes
both works far away from the optimal solution.  To optimize the system performance (both SE and EE) of the MIMO BC, the precoding matrices
and power allocation should be jointly decided. For instance, the common way to achieve the maximum sum capacity is employing the dirty paper
coding (DPC) and iterative waterfilling \cite{DPC,Duality,IterativeBC,IterativeMAC}. From the standpoint of EE, joint precoding and power
allocation design should also be carefully designed to achieve the optimum, which is the main concern of this paper.

We consider the problem of optimizing EE with joint precoding and power allocation for the MIMO BC, where a practical power model including
signal processing, circuit power, etc. at the base station (BS) \cite{Xu1,Arnold} is taken into account. After formulating the maximum EE problem, we find that the energy efficient joint precoding and power allocation
design is equivalent to only optimizing the transmit covariance matrices. However, optimizing the transmit covariance matrices is difficult, as
the EE function  achieved by DPC is nonconcave. Fortunately, we find an efficient iterative solution based on uplink-downlink duality and the
contributions are summarized as follows.


{\emph{Contributions:}} We transform the EE into a quasiconcave function through employing the famous uplink-downlink duality. The duality
transforms the nonconcave MIMO BC capacity into the dual convex MIMO multi-access channels (MAC). After that, we propose a novel energy
efficient iterative waterfilling scheme based on the block-coordinate ascent algorithm to solve the quasiconcave EE optimizing problem
efficiently. During each iteration, the transmit covariance matrices optimization is formulated as a concave fractional program, and solved
through relating it to a parametric concave program and then applying the Karush-Kuhn-Tucker optimality conditions. Interestingly, the solution
of each iteration has a feature of waterfilling. We prove the convergence of the proposed scheme and validates it through simulations. Moreover,
the system parameters' effect on the  EE is discussed finally.


%


\section{System Model}

\begin{figure}[t]
\begin{center}
\includegraphics[height = 2in] {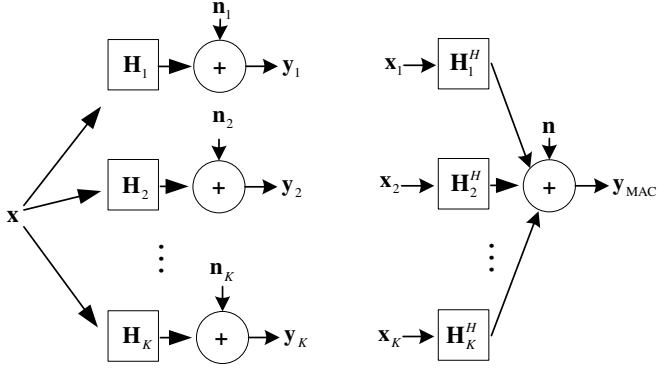}
\end{center}
\caption{System model of the MIMO BC (left) and its dual MIMO MAC (right).} \label{fig000}
\end{figure}

The system consists of a single BS with $M$ antennas and $K$ users each with $N$ antennas{\footnote{The results here can be extended to the
general case with different antenna number at each user. Moreover, the results are also applicable to the multi-cell scenario with BS
cooperation.}}, which is shown in Fig. \ref{fig000}. The downlink channel can be denoted as
\begin{equation} \label{eq1}
\begin{array}{l}
{\bf{y}}_i = {\bf{H}}_i {\bf{x}} + {\bf{n}}_i, i=1,\ldots,K,
\end{array}
\end{equation}
and the dual uplink channel is denoted as
\begin{equation} \label{eq2}
\begin{array}{l}
{\bf{y}}_{\rm{MAC}} = \sum \limits _{i=1}^{K}{\bf{H}}_i^{H} {\bf{x}}_i + {\bf{n}},
\end{array}
\end{equation}
where ${\bf{H}}_i \in {\mathbb{C}}^{N \times M}$ is the channel matrix of the $i$th user, ${\bf{x}} \in {\mathbb{C}}^{M \times 1}$ is the
transmitted signal on the downlink, ${\bf{x}}_i\in {\mathbb{C}}^{N \times 1}$ is the transmitted signal on the uplink, ${\bf{n}}_i\in
{\mathbb{C}}^{N \times 1}$ and ${\bf{n}}\in {\mathbb{C}}^{M \times 1}$ are the independent Gaussian noise with each entry
${\mathcal{CN}}(0,\sigma^2)$. Frequency flat fading channels with bandwidth $W$ is considered and the channel state information (CSI) is assumed
to be perfectly known at the transmitter and receivers.

About the power model, as BSs take the main power consumption in the
cellular networks, the users' consumed power is omitted. The BS
power model is motivated by \cite{Xu1,Arnold}. As the power radiated
to the environment for signal transmission is only a portion of its
total power consumption \cite{Arnold}, the practical circuit power,
signal processing power, cooling loss etc. at the BS should be taken
into account. Without loss of generality, given the total BS antenna
number $M$ and total transmit power $P$, the total power consumption
of a BS can be denoted as
\begin{equation} \label{eq30}
\begin{array}{l}
\displaystyle P_{\rm{total}} = \frac{P}{\eta} + M P_{\rm{dyn}} + P_{\rm{sta}},
\end{array}
\end{equation}
where $\eta$ denotes the power amplifier (PA) efficiency; $M P_{\rm{dyn}}$ denotes the dynamic power consumption proportional to the number of
radio frequency (RF) chains, e.g. circuit power of RF chains which is always proportional to $M$; and $P_{\rm{sta}}$ accounts for the static
power independent of both $M$ and $P$ which includes power consumption of the baseband processing, battery unit etc..

\section{Achievable Sum Capacity and Energy Efficiency}

The sum capacity of the MIMO BC is achieved by DPC, which can be denoted as follows \cite{DPC} given a total transmit power $P$.
\begin{equation} \label{eq3}\begin{array}{l}
\displaystyle {C_{{\rm{BC}}}}\left( {{{\bf{H}}_1}, \ldots ,{{\bf{H}}_K},P} \right)\\
\displaystyle = \mathop {\max }\limits_{\left\{ {{{\bf{\Sigma }}_i}} \right\}_{i = 1}^K:{{\bf{\Sigma }}_i} \ge 0,\sum\nolimits_{i = 1}^K {{\rm{Tr}}\left( {{{\bf{\Sigma }}_i}} \right)}  \le P} W\log \left| {{\bf{I}} + \frac{1}{\sigma^2}{{\bf{H}}_1}{{\bf{\Sigma }}_1}{\bf{H}}_1^H} \right|\\
 \displaystyle+W \log \frac{{\left| {{\bf{I}} + \frac{1}{\sigma^2}{{\bf{H}}_2}\left( {{{\bf{\Sigma }}_1} + {{\bf{\Sigma }}_2}} \right){\bf{H}}_2^H} \right|}}{{\left| {{\bf{I}} + \frac{1}{\sigma^2}{{\bf{H}}_2}\left( {{{\bf{\Sigma }}_1}} \right){\bf{H}}_2^H} \right|}} +  \cdots \\
\displaystyle + W\log \frac{{\left| {{\bf{I}} + \frac{1}{\sigma^2}{{\bf{H}}_K}\left( {{{\bf{\Sigma }}_1} +  \cdots  + {{\bf{\Sigma }}_K}}
\right){\bf{H}}_K^H} \right|}}{{\left| {{\bf{I}} + \frac{1}{\sigma^2}{{\bf{H}}_K}\left( {{{\bf{\Sigma }}_1} +  \cdots  + {{\bf{\Sigma }}_{K -
1}}} \right){\bf{H}}_K^H} \right|}},
\end{array}
\end{equation}
where the optimization is performed to choose the optimal downlink transmit covariance matrices ${{\bf{\Sigma }}_i} \in \mathbb{C}^{M \times M},
i=1,\ldots,K$.

The EE is defined as the achievable sum capacity of MIMO BC divided by the total power consumption at the BS, which can be denoted as 
\begin{equation} \label{eq301}\begin{array}{l}
\displaystyle  \xi_{\rm{BC}} \left( {{{\bf{H}}_1}, \ldots ,{{\bf{H}}_K}}, P\right) =  \frac{ \displaystyle
{C_{{\rm{BC}}}}\left( {{{\bf{H}}_1}, \ldots ,{{\bf{H}}_K},P} \right)}{\frac{P}{\eta} + M P_{\rm{dyn}} + P_{\rm{sta}}}
\end{array}
\end{equation}
under a fixed transmit power $P$. Based on (\ref{eq301}), the optimal EE for the MIMO BC can be obtained through optimizing $P${\footnote{Indeed, the EE of MIMO BC is highly affected by the transmit power $P$ and transmit antenna number $M$, and jointly optimizing $P$ and $M$ is required to maximize the EE. We only consider the optimization of $P$ in this paper, and the $M$ optimizing to further improve the EE based on active transmit antenna selection can be found in \cite{XuTWC}.}}, which is denoted as
\begin{equation} \label{eq4}\begin{array}{l}
\displaystyle  \xi_{\rm{BC}} \left( {{{\bf{H}}_1}, \ldots ,{{\bf{H}}_K}}\right) = \max \limits_{P:P\geq0}  \frac{ \displaystyle
{C_{{\rm{BC}}}}\left( {{{\bf{H}}_1}, \ldots ,{{\bf{H}}_K},P} \right)}{\frac{P}{\eta} + M P_{\rm{dyn}} + P_{\rm{sta}}}.
\end{array}
\end{equation}
In the above optimization problem, transmit power level $P$ and transmit covariance matrices ${{\bf{\Sigma }}_i} \in \mathbb{C}^{M \times M},
i=1,\ldots,K$ need to be jointly optimized. Note that although no maximum transmit power constraint is considered in (\ref{eq4}), the solution
can be easily extended to the constrained case based on \cite{Miao3,PowerMin}. We omit the extension here due to page limit.

As (\ref{eq4}) is nonconcave,  optimizing (\ref{eq4}) is nontrivial. Fortunately, motivated by \cite{Chong3,frac_program}, we find out that the
following property. If the numerator (sum capacity) can be transformed into a convex function, the EE can be formulated as a quasiconcave
function, because the denominator (total power consumption) is affine. Based on this observation, we try to transform the sum capacity into a
concave function.

Applying the uplink-downlink duality \cite{Duality}, the MIMO BC sum capacity (\ref{eq3}) is equal to the concave sum capacity of the MIMO MAC
(\ref{eq2}) with sum transmit power constraint, which can be denoted as
\begin{equation} \label{eq5}\begin{array}{l}
\displaystyle {C_{{\rm{MAC}}}}\left( {{{\bf{H}}_1^H}, \ldots ,{{\bf{H}}_K^H},P} \right) \\
\displaystyle =\mathop {\max }\limits_{\left\{ {{{\bf{Q}}_i}} \right\}_{i = 1}^K:{{\bf{Q}}_i} \ge 0,\sum\nolimits_{i = 1}^K {{\rm{Tr}}\left(
{{{\bf{Q}}_i}} \right)}  \le P} W\log \left| {{\bf{I}} + \frac{1}{\sigma^2}\sum\limits_{i = 1}^K {{\bf{H}}_i^H{{\bf{Q}}_i}{{\bf{H}}_i}} }
\right|,
\end{array}
\end{equation}
where the uplink transmit covariance matrices ${{\bf{Q }}_i} \in \mathbb{C}^{N \times N}, i=1,\ldots,K$ need to be optimized. Thus, the dual MAC
optimal EE should be rewritten as
\begin{equation} \label{eq6}\begin{array}{l}
\displaystyle  \xi_{\rm{MAC}} \left( {{{\bf{H}}_1}, \ldots ,{{\bf{H}}_K}}\right) = \max \limits_{P:P\geq0}  \frac{ \displaystyle
{C_{{\rm{MAC}}}}\left( {{{\bf{H}}_1^H}, \ldots ,{{\bf{H}}_K^H},P} \right)}{\frac{P}{\eta} + M P_{\rm{dyn}} + P_{\rm{con}}}.
\end{array}
\end{equation}
According to the duality and the mapping between ${\bf{\Sigma}}_i$ and ${\bf{Q}}_i$ \cite{Duality}, optimal ${\bf{\Sigma}}_i, i=1,\ldots,K$ and
$P$ can be obtained if we can get the optimal ${\bf{Q}}_i, i=1,\ldots,K$ and $P$ in (\ref{eq6}). Furthermore, since the maximum power constraint
is not considered and $\sum\limits_{i = 1}^K {{\rm{Tr}}\left( {{{\bf{Q}}_i}} \right)} = P$ is always required for optimizing (\ref{eq5})
\cite{IterativeBC}, (\ref{eq6}) can be simplified and rewritten as
\begin{equation} \label{eq7}\begin{array}{l}
\displaystyle {\xi_{{\rm{MAC}}}}\left( {{{\bf{H}}_1^H}, \ldots ,{{\bf{H}}_K^H}} \right) \\
\displaystyle =\mathop {\max }\limits_{\left\{ {{{\bf{Q}}_i}} \right\}_{i = 1}^K:{{\bf{Q}}_i} \ge 0}\frac{W \log \left| {{\bf{I}} +
\frac{1}{\sigma^2}\sum\limits_{i = 1}^K {{\bf{H}}_i^H{{\bf{Q}}_i}{{\bf{H}}_i}} } \right|}{\frac{{\sum\nolimits_{i = 1}^K{{\rm{Tr}}\left(
{{{\bf{Q}}_i}} \right)}}}{\eta} + M P_{\rm{dyn}} + P_{\rm{sta}}}.
\end{array}
\end{equation}
Finally, the optimization of (\ref{eq4}) is transformed into optimizing (\ref{eq7}). If we can obtain the optimal ${\bf{Q}}_i, i=1,\ldots,K$ for
(\ref{eq7}), the optimal $P$ can be decided correspondingly based on $\sum\limits_{i = 1}^K {{\rm{Tr}}\left( {{{\bf{Q}}_i}} \right)} = P$. That
is to say, designing the joint precoding and power allocation to maximize EE is equivalent to optimizing the transmit covariance matrices. Once
transmit covariance matrices are decided, the transmit power level can be correspondingly determined.

Let us look at (\ref{eq7}) again. Since the numerator is concave and the denominator is affine, (\ref{eq7}) is a quasiconcave optimization,
which can be solved through the bisection method or interior-point methods. However, the numerical methods would be still too complex when the
user number becomes significantly large. Motivated by \cite{IterativeBC,IterativeMAC}, an energy efficient iterative waterfilling is proposed in
the next section to solve it more efficiently.

\section{Energy Efficient Iterative Optimization}

\subsection{Motivation}\label{sec3a}

As the EE is distinct from the capacity, the spectral efficient iterative waterfilling \cite{IterativeBC} is not applicable for the EE any
longer. Nevertheless, we notice that the basic idea of the spectral efficient iterative algorithms are based on the block-coordinate ascent
algorithm \cite[Sec. 2.7]{NonlinearProgramming}. That is to say, if we can write the EE as the similar structure with the block-coordinate
ascent algorithm and then prove it satisfies the condition of \cite[Sec. 2.7]{NonlinearProgramming}, we can obtain an iterative solution of the
problem (\ref{eq7}).

For ease of description, we define the following function $g \left( \cdot \right)$ at first.
\begin{equation} \label{eq8}\begin{array}{l}
g \left( {{{\bf{Q}}_1}, \ldots ,{{\bf{Q}}_K}} \right) = \frac{W \log \left| {{\bf{I}} + \frac{1}{\sigma^2}\sum\limits_{i = 1}^K
{{\bf{H}}_i^H{{\bf{Q}}_i}{{\bf{H}}_i}} } \right|}{\frac{{\sum\nolimits_{i = 1}^K{{\rm{Tr}}\left( {{{\bf{Q}}_i}} \right)}}}{\eta} + M
P_{\rm{dyn}} + P_{\rm{sta}}}.
\end{array}
\end{equation}
For the block-coordinate ascent algorithm, given the current iterate ${{\bf{Q}}^{(k)}} = \left({{{\bf{Q}}_1^{(k)}}, \ldots
,{{\bf{Q}}_K^{(k)}}}\right)$, the next iterate ${{\bf{Q}}^{(k+1)}} = \left({{{\bf{Q}}_1^{(k+1)}}, \ldots ,{{\bf{Q}}_K^{(k+1)}}}\right)$ can be
generated as
\begin{equation} \label{eq801}\begin{array}{l}
{{\bf{Q}}_i^{(k+1)}} \\ =\displaystyle \arg \max \limits_{{{\bf{Q}}_i}:{\bf{Q}}_i \ge 0} g \left( {{{\bf{Q}}_1^{(k+1)}}, \ldots
,{{\bf{Q}}_{i-1}^{(k+1)}},
 {{\bf{Q}}_i},{{\bf{Q}}_{i+1}^{(k)}},\ldots,{{\bf{Q}}_K^{(k)}}}
 \right).
\end{array}
\end{equation}
However, to apply the iterative algorithm efficiently, there are
conditions need to be satisfied. For one thing, the solution of
({\ref{eq801}}) should be uniquely attained \cite[Proposition
2.7.1]{NonlinearProgramming}. For another, the solution should be
simple and easy to employ.

Very fortunately, the two conditions both fulfill and the solution can be obtained following an energy efficient waterfilling feature. We are
interested to show it in the next subsection.

%
%
%
%
%
%

\subsection{Energy Efficient Waterfilling} \label{EEW}

Based on \cite{IterativeBC,IterativeMAC}, it is fulfilled that
\begin{equation} \label{eq9}
\begin{array}{l}
\log \left| {{\bf{I}} +
\frac{1}{\sigma^2}\sum\limits_{i = 1}^K {{\bf{H}}_i^H{{\bf{Q}}_i}{{\bf{H}}_i}} } \right| \\
= \log \left| {{\bf{I}} + \frac{1}{\sigma^2}\sum\limits_{j \ne i} {{\bf{H}}_j^H{{\bf{Q}}_j}{{\bf{H}}_j}} } \right|  \\+ \log \left| {\bf{I}} +
{{\left( \sigma^2{{\bf{I}} + \sum\limits_{j \ne i} {{\bf{H}}_j^H{{\bf{Q}}_j}{{\bf{H}}_j}} } \right)}^{ - 1/2}} \right.  \\ \left.\times
{\bf{H}}_i^H{{\bf{Q}}_i}{{\bf{H}}_i}{{\left( \sigma^2{{\bf{I}} + \sum\limits_{j \ne i} {{\bf{H}}_j^H{{\bf{Q}}_j}{{\bf{H}}_j}} } \right)}^{ -
1/2}}\right| \\
 = \log \left| {{{\bf{Z}}_i}} \right| + \log \left| {{\bf{I}} + {\bf{G}}_i^H{{\bf{Q}}_i}{{\bf{G}}_i}} \right|,
\end{array}
\end{equation}
where ${{\bf{Z}}_i} = {\bf{I}} + \frac{1}{\sigma^2}\sum\limits_{j \ne i} {{\bf{H}}_j^H{{\bf{Q}}_j}{{\bf{H}}_j}}$ and ${{\bf{G}}_i} =
{{\bf{H}}_i}{\left( \sigma^2{{\bf{I}} + \sum\limits_{j \ne i} {{\bf{H}}_j^H{{\bf{Q}}_j}{{\bf{H}}_j}} } \right)^{ - 1/2}}$. By denoting
\[a_i = \frac{ \sum\limits_{j \neq i
 }{{\rm{Tr}}\left( {{{\bf{Q}}_j}} \right)}}{\eta} + M P_{\rm{dyn}} +
P_{\rm{sta}},\]
\[
b_i = W \log \left| {{{\bf{Z}}_i}} \right|
\]
and substituting (\ref{eq9}) into (\ref{eq8}) we have that
\begin{equation} \label{eq10}\begin{array}{l}
g \left( {{{\bf{Q}}_1}, \ldots ,{{\bf{Q}}_K}} \right) = \displaystyle \frac{b_i + W\log \left| {{\bf{I}} + {\bf{G}}_i^H{{\bf{Q}}_i}{{\bf{G}}_i}}
\right|}{\frac{{\rm{Tr}}\left( {{{\bf{Q}}_i}} \right)}{\eta} + a_i}
\end{array}
\end{equation}

%

Therefore, we can redefine the problem (\ref{eq801}) by removing the
iteration number as to
\begin{equation} \label{eq12}
\begin{array}{l}
\mathop {{\rm{maximize}}}\limits_{{{\bf{Q}}_i}:{\bf{Q}}_i \ge 0} g \left( {{{\bf{Q}}_1}, \ldots
,{{\bf{Q}}_{i-1}},{{\bf{Q}}_i},{{\bf{Q}}_{i+1}},\ldots,{{\bf{Q}}_K}} \right) \\
= \displaystyle \frac{b_i + W\log \left| {{\bf{I}} +
{\bf{G}}_i^H{{\bf{Q}}_i}{{\bf{G}}_i}} \right|}{\frac{{\rm{Tr}}\left(
{{{\bf{Q}}_i}} \right)}{\eta} + a_i}
\end{array}
\end{equation}
by treating ${{{\bf{Q}}_1}, \ldots ,{{\bf{Q}}_{i-1}},{{\bf{Q}}_{i+1}},\ldots,{{\bf{Q}}_K}} $ as constant. Based on section \ref{sec3a}, we need
to solve the above problem and prove that the solution is unique.

Since the numerator and denominator in (\ref{eq10}) are concave and
affine respectively, (\ref{eq12}) is a concave fractional program
\cite{frac_program}. Define a non-negative parameter $\lambda$,
(\ref{eq12}) is related to the following convex function separating
numerator and denominator with help of $\lambda$.
\begin{equation} \label{eq13}
\begin{array}{l}
F({{\bf{Q}}_i},\lambda) = b_i + W\log \left| {{\bf{I}} + {\bf{G}}_i^H{{\bf{Q}}_i}{{\bf{G}}_i}} \right| -\lambda  \left( {{\frac{{{\rm{Tr}}\left(
{{{\bf{Q}}_i}} \right)}}{\eta} + a_i}}\right)
\end{array}
\end{equation}
And then define a convex optimization problem as
\begin{equation} \label{eq13001}
\begin{array}{l}
Y(\lambda) = \max \limits _{{{\bf{Q}}_i}:{{\bf{Q}}_i} \ge 0} F({{\bf{Q}}_i},\lambda).
\end{array}
\end{equation}
We will try to solve (\ref{eq13001}), and then the solution of (\ref{eq12}) can be obtained correspondingly based on the following Theorem.

\newtheorem{Theorem}{Theorem}
\begin{Theorem}\label{Theorem0}
The optimum feasible transmit covariance matrix ${\bf{Q}}_i^*$ achieves the maximum value of (\ref{eq12}) if and only if $Y(\lambda^*) =
F({{\bf{Q}}_i^*},\lambda^*) = \max F({{\bf{Q}}_i},\lambda^*|{{\bf{Q}}_i} \geq 0) = 0$.
\end{Theorem}

{\emph{Proof}}: See Appendix \ref{appB}.

Theorem \ref{Theorem0} gives us insights to solve (\ref{eq12}). We should optimizing (\ref{eq13001}) at first under a given $\lambda$ and then
solve the equation $Y(\lambda) = 0$ to get the optimal $\lambda$.

To solve (\ref{eq13001}), we can denote
\begin{equation} \label{eq1301}
\begin{array}{l}
{\bf{G}}_i^H{{\bf{G}}_i} = {\bf{U}}{\bf{D}}_i{\bf{U}}^H
\end{array}
\end{equation}
based on the eigenvalue decomposition at first, where ${\bf{D}}_i \in{\mathbb C}^{M \times M}$ is diagonal with nonnegative entries and
${\bf{U}} \in{\mathbb C}^{M \times M}$ is unitary. Without loss of generality, we assume that ${\bf{D}}_i$ has $L$ non-zero diagonal entries ($1 \leq L \leq M$), which
means $[{\bf{D}}_i]_{kk} > 0$ for $k = 1,\ldots,L$ and $[{\bf{D}}_i]_{kk} = 0$ for $k = L+1,\ldots,M$.

And then we have the following equation based on ${\bf{I+AB}} = {\bf{I+BA}}$ \cite{IterativeBC}:
\begin{equation} \label{eq16}
\begin{array}{l}
\log \left| {{\bf{I}} + {\bf{G}}_i^H{{\bf{Q}}_i}{{\bf{G}}_i}} \right| = \log \left| {{\bf{I}} + {{\bf{Q}}_i}{\bf{G}}_i^H{{\bf{G}}_i}} \right|
\\ = \log \left| {{\bf{I}} + {{\bf{Q}}_i}{\bf{U}}{\bf{D}}_i{\bf{U}}^H} \right| = \log \left| {{\bf{I}} + {\bf{U}}^H{{\bf{Q}}_i}{\bf{U}}{\bf{D}}_i}
\right|
\end{array}
\end{equation}
Define ${\bf{S}}_i = {\bf{U}}^H{{\bf{Q}}_i}{\bf{U}}$. As ${\bf{U}}$ is unitary, we have that ${\rm{Tr}}({\bf{S}}_i) = {\rm{Tr}}({\bf{Q}}_i)$.
Thus, (\ref{eq13}) can be rewritten as
\begin{equation} \label{eq17}
\begin{array}{l}
G({\bf{S}}_i,\lambda)=     { b_i + W\log \left| {{\bf{I}} + {\bf{S}}_i{{\bf{D}}}_i} \right|} -\lambda \left({{\displaystyle
 \frac{{{\rm{Tr}}\left( {{{\bf{S}}}_i} \right)} }{\eta} + a_i }}\right)
\end{array}
\end{equation}

As each ${\bf{S}}_i$ corresponds to a ${\bf{Q}}_i$ via the
invertible mapping ${\bf{S}}_i = {\bf{U}}^H{{\bf{Q}}_i}{\bf{U}}$,
solving (\ref{eq13001}) is equivalent to solving the following
convex optimization problem.
\begin{equation} \label{eq18}
\begin{array}{l}
Y(\lambda) = \max \limits _{{{\bf{S}}_i}:{{\bf{S}}_i} \ge 0}
G({{\bf{S}}_i},\lambda)
\end{array}
\end{equation}

It is proved in the Appendix \ref{appendix1} that the optimal ${{\bf{S}}_i^*}$ to solve (\ref{eq18}) is diagonal with $[{\bf{S}}_i^*]_{kk} > 0$
for $k = 1,\ldots,L$ and $[{\bf{S}}_i^*]_{kk} = 0$ for $k = L+1,\ldots,M$. Thus, $G({\bf{S}}_i,\lambda)$ with diagonal ${\bf{S}}_i$ is
\begin{equation} \label{eq19}
\begin{array}{l}
G({\bf{S}}_i,\lambda)=     b_i + W \sum \limits _{k=1}^{L} \log \left( {{{1}} + [{\bf{S}}_i]_{kk}[{{\bf{D}}}_i}]_{kk} \right) \\ -\lambda
\left({{
 \frac{\sum \limits _{k=1}^ L{ {[{{\bf{S}}}_i]_{kk}} } }{\eta} + a_i }}\right).
\end{array}
\end{equation}
As (\ref{eq19}) is concave in ${{\bf{S}}_i}$, the problem (\ref{eq18}) can be solved for a given $\lambda$ by solving the Karush-Kuhn-Tucker
optimality conditions, and the solution can be denoted as
\begin{equation} \label{eq20}
\begin{array}{l}
[{{\bf{S}}_{i}^*}]_{kk}^{\lambda} = \left[\frac{{\eta  }}{{\ln(2)\lambda}} - \frac{1}{{[{{\bf{D}}_{i}}]_{kk}}}\right]^+,  k = 1,\ldots,L,
\end{array}
\end{equation}
where $[x]^+ = \max(x,0)$. Then the water level $\lambda^*$ can be decided by setting $Y(\lambda^*) = 0$ based on Theorem \ref{Theorem0} as
\begin{equation} \label{eq21}
\begin{array}{l}
{ b_i + \sum \limits _{k=1}^{L} \log \left( {{{1}} + \left[\frac{{\eta  }}{{\ln(2)\lambda^*}} -
\frac{1}{{[{{\bf{D}}_{i}}]_{kk}}}\right]^+[{{\bf{D}}}_i}]_{kk} \right)}
 \\-\lambda^* \times \left(
 \frac{\sum \limits _{k=1}^ L{ {\left[\frac{{\eta  }}{{\ln(2)\lambda^*}} - \frac{1}{{[{{\bf{D}}_{i}}]_{kk}}}\right]^+} } }{\eta} + a_i \right) = 0.
\end{array}
\end{equation}
As $Y(\lambda)$ is strictly decreasing, and $F(0) = \infty$, $F(\infty)=-\infty$  (see detailed proof in Appendix \ref{AppC}), we can solve (\ref{eq21}) efficiently based on the bisection methods.

Based on (\ref{eq21}) and (\ref{eq20}), the optimal
${{\bf{S}}_i^*}^{\lambda ^*}$ is derived. Based on the mapping
between ${{\bf{S}}_i}$ and ${{\bf{Q}}_i}$, finally, the optimal
solution of (\ref{eq12}) can be derived as
\begin{equation} \label{eq22}
\begin{array}{l}{{\bf{Q}}_i^*} = {\bf{U}}{{\bf{S}}_i^*}^{\lambda ^*}{\bf{U}}^H .
\end{array}
\end{equation}

To prove that the solution of (\ref{eq12}) is unique, we only need to prove that $\lambda^*$ is unique. We give the following Theorem and the
proof is given in the Appendix \ref{AppC}.
\begin{Theorem}\label{Theorem1}
The derived water level $\lambda^*$ in (\ref{eq21}) is unique and globally optimal.
\end{Theorem}

To make the description more clearly, we summarize the energy efficient waterfilling algorithm for optimizing (\ref{eq10}) in TABLE
\ref{TABLE0}.

\begin{table}[!h]
\renewcommand{\arraystretch}{1.3}
\caption{Energy Efficient Waterfilling Algorithm} \label{TABLE0} \centering
\begin{tabular}{|p{3.2in}|}
\hline
\begin{enumerate}
\item Calculate ${{\bf{Z}}_i} = {\bf{I}} + \frac{1}{\sigma^2}\sum\limits_{j \ne i} {{\bf{H}}_j^H{{\bf{Q}}_j}{{\bf{H}}_j}}$,
${{\bf{G}}_i} = {{\bf{H}}_i}{\left( \sigma^2{{\bf{I}} + \sum\limits_{j \ne i} {{\bf{H}}_j^H{{\bf{Q}}_j}{{\bf{H}}_j}} } \right)^{ - 1/2}}$, $a_i
= \frac{ \sum\limits_{j \neq i
 }{{\rm{Tr}}\left( {{{\bf{Q}}_j}} \right)}}{\eta} + M P_{\rm{dyn}} +
P_{\rm{sta}}$, $ b_i = W \log \left| {{{\bf{Z}}_i}} \right|; $
\item Define the related parametric convex program in (\ref{eq13}) and (\ref{eq13001});
\item Transform the parametric convex program into diagonal forms (\ref{eq17}) and (\ref{eq18}) by performing
 eigenvalue decomposition in $(\ref{eq1301})$;
\item Solve (\ref{eq18}) by solving the Karush-Kuhn-Tucker
optimality conditions and obtain the solution ${{\bf{S}}_i^*}^{\lambda}$ in (\ref{eq20});
\item Calculate the energy efficient water level ${\lambda ^*}$ based on (\ref{eq21}) and determine the optimal ${{\bf{S}}_i^*}^{\lambda ^*}$;
\item Obtain ${{\bf{Q}}_i^*}$ based on the mapping (\ref{eq22}) finally;
\end{enumerate}
\\
\hline
\end{tabular}
\end{table}

\subsection{Iterative Algorithm}

Based on the derivation in section \ref{EEW} and the
block-coordinate ascent algorithm, the energy efficient iterative
waterfilling scheme can be derived as shown in TABLE \ref{table_1},
and the proof of converge is given as follows.

\begin{table}[!h]
\renewcommand{\arraystretch}{1.3}
\caption{Energy Efficient Iterative waterfilling Scheme} \label{table_1} \centering
\begin{tabular}{|p{3.2in}|}
\hline
\textbf{Initialization}: Set ${{\bf{Q}}_i} = {\bf{0}}, i = 1,\ldots,K.$,\\
\textbf{Repeat}:\\
\quad For $i=1:K$
\begin{enumerate}
\item Calculate ${{\bf{Q}}_i^*}$ based on the energy efficient waterfilling algorithm in TABLE \ref{TABLE0};
\item Refresh ${\bf{Q}}_i$ as ${{\bf{Q}}_i^*}$;
\end{enumerate}
\quad End \\
\textbf{Until the EE converges.}\\
\hline
\end{tabular}
\end{table}

\emph{Proof of converge}: Firstly, during each step, the energy
efficient waterfilling can achieve an global maximization of (\ref{eq12}) treating the
other users' transmit covariance matrices as constant, the EE is
non-decreasing with each step. As the EE is bounded, the EE
converges to a limit.

Secondly, according to Theorem \ref{Theorem0} and Theorem
\ref{Theorem1}, the derivation of each step is unique. Based on
\cite[Sec. 2.7]{NonlinearProgramming}, the set of
${\bf{Q}}_1,\ldots,{\bf{Q}}_K$ also converge to a limit. $\Box$

Note that as the proof does not depend on the starting point, we can start the algorithm from any starting values of
${\bf{Q}}_1,\ldots,{\bf{Q}}_K$. To show the efficiency of the proposed scheme, we give the simulation results in the next section.

\section{Simulation Results}

\begin{figure}[t]
\begin{center}
\includegraphics[height = 2.4in] {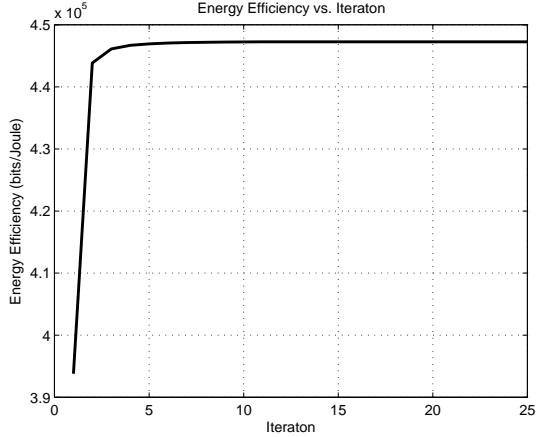}
\end{center}
\caption{EE converge behavior of the proposed scheme.} \label{fig001}
\end{figure}

\begin{figure}[t]
\begin{center}
\includegraphics[height = 2.4in] {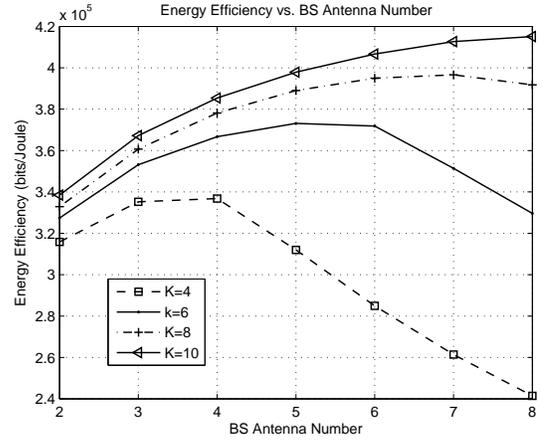}
\end{center}
\caption{The effect of antenna number $M$ on the EE, where $N=1$, $d=$1km are considered.} \label{fig002}
\end{figure}

\begin{figure}[t]
\begin{center}
\includegraphics[height = 2.4in] {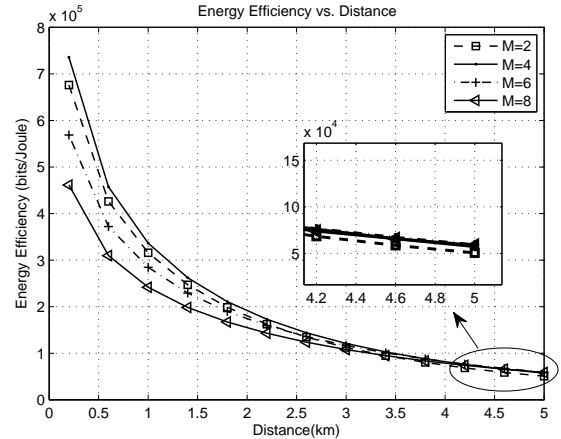}
\end{center}
\caption{Effect of distance between BS and users on the EE, where $N=1$, $K=4$ are considered.} \label{fig003}
\end{figure}

\begin{table}[t]
\caption{Simulation Parameters} \label{table:1}
\begin{center}
\begin{tabular}{|p{2.5cm}|p{3cm}|}
\hline Bandwidth & 5MHz \\
\hline Noise power & -110dBm \\
\hline Pathloss & $128.1+37.6 \log_{10}d_{i,j}$ ($d$ in kilometers) \\
\hline $P_{\rm{dyn}}$ & 83W\\
\hline $P_{\rm{Sta}}$ & 45.5W\\
\hline $\eta$ & 0.38\\
 \hline
\end{tabular}
\end{center}
\end{table}
In the simulation, large scale pathloss and small scale Rayleigh fading are considered. The parameters are set in TABLE \ref{table:1} based on
\cite{Xu1} and all users are with the same distance. In Fig. \ref{fig001}, the converge behavior of the proposed scheme is shown and it is set
that $d=1$km, $M=4$, $N=4$ and $K=10$. We can see that our proposed iterative scheme converges very fast. It can achieve the optimal EE under
nearly five iterations.

Fig. \ref{fig002} compares the effect of $M$ on the EE with
different $K$, where $N=1$, $d=$1km are considered. Different from
the SE which is always increasing linearly as $M$ increases when $M
\leq N\times K$, the EE increases much slower or even decreases. For
example, when $K=8$, the EE with $M=7$ is better than $M=8$. The
reason comes from the effect of the practical dynamic power $M
P_{\rm{dyn}}$. In this case, increasing $M$ can cause the linear
increasing of both capacity and dynamic power. Moreover, when $M
> K\times N$, the increasing of $M$ always decreases the EE
performance. That is because the benefits of transmit diversity gain
caused by the increasing of $M$ is much smaller than the drawbacks
of dynamic power increasing. The user number $K$ and transmit
antenna number $M$ affect the EE in a complicated manner, adjusting
these parameters adaptively is important for improving the EE. This
is distinct from the spectral efficient systems, where more $M$
always benefits.

Fig. \ref{fig003} compares the effect of $d$ on the EE with
different $M$, where $N=1$, $K=4$ are considered. It is observed
that the optimal BS antenna number is different under different $d$.
For example, $M=4$ is optimal when $d=0.2$km and $M=2$ is less best,
but the trends change when $K=5$km, where $M=2$ performs worst. This
situation varies due to the tradeoff among capacity, transmit power,
dynamic power and static power. For instance, when the distance is
large, transmit power would take the main part of the total power
consumption, thus, $M=2$ with the smallest BS antennas number would
consume highest transmit power and has the  worst EE. How to study
these parameters in a comprehensive manner is a challenge, which
should be left for the future work.

\section{Conclusion}

Based on uplink-downlink duality, the EE of the MIMO BC can be transformed into a quasiconcave problem. Based on this feature, we propose an
energy efficient iterative waterfilling scheme to maximize the EE for the MIMO BC based on the block-coordinate ascent algorithm. We prove the
converge of the proposed scheme and validates it through simulations. Finally, the effect of system parameters is discussed.

 \appendices

\section{}\label{appB} 
As in (\ref{eq12}) the numerator is concave and differentiable, and the denominator is convex and differentiable, Theorem \ref{Theorem0} can be
directly obtained based on \cite[Proposition 6]{frac_program}.

\section{} \label{appendix1}
The proof is motivated by \cite[Appendix II]{IterativeBC}.

We prove that $[{\bf{S}}_i^*]_{jk} = 0, \forall j,k >L$ at first. Consider ${\bf{S}}\geq 0$ with $[{\bf{S}}]_{jk} \ne 0$ for some $j>L$ and
$k>L$. Based on \cite[Appendix II]{IterativeBC}, we have $\sum \nolimits _{k=L+1}^M [{\bf{S}}]_{kk} > 0$. Thus, we can redefine another diagonal
matrix ${\bf{S}}' \geq 0$ as
\begin{equation} \label{A1}
\begin{array}{l}
{\left[ {{{\bf{S}}^\prime }} \right]_{kk}} = \left\{ {\begin{array}{*{20}{c}}
{{{\left[ {\bf{S}} \right]}_{11}} + \sum\limits_{j = L + 1}^M {{{\left[ {\bf{S}} \right]}_{jj}}},\quad k = 1}\\
{{{\left[ {\bf{S}} \right]}_{kk}},\quad k = 2, \ldots ,L}\\
{0,\quad  k = L + 1, \ldots ,M}
\end{array}} \right.
\end{array}
\end{equation}
with ${\rm{Tr}}({\bf{S}}') = {\rm{Tr}}({\bf{S}})$ and $\log |{\bf{I+S'D}}_i| > \log |{\bf{I+SD}}_i|$. Hence, $G({\bf{S}}',\lambda) >
G({\bf{S}},\lambda)$. Therefore, $[{\bf{S}}_i^*]_{jk} = 0, \forall j,k >L$.

We need to prove that ${\bf{S}}_i^*$ is diagonal then. Consider any ${\bf{S}}\geq 0$ with $[{\bf{S}}]_{jk} = 0$ for any $j>L$ and $k>L$ but is
not diagonal. We can have another diagonal matrix ${\bf{S}}'$ with $[{\bf{S}}']_{kk} = [{\bf{S}}]_{kk}, k=1,\ldots,M$. Based on \cite[Appendix
II]{IterativeBC}, ${\rm{Tr}}({\bf{S}}') = {\rm{Tr}}({\bf{S}})$ and $\log |{\bf{I+S'D}}_i| > \log |{\bf{I+SD}}_i|$.

Therefore, the optimal ${{\bf{S}}_i^*}$ is diagonal with $[{\bf{S}}_i^*]_{kk} > 0$ for $k = 1,\ldots,L$ and $[{\bf{S}}_i^*]_{kk} = 0$ for $k =
L+1,\ldots,M$.

\section{}\label{AppC}

As in (\ref{eq12}) the numerator is concave and continuous, and the denominator is convex and continuous, $F(\lambda)$ is strictly decreasing
and continuous based on \cite{frac_program}.

Look at ({\ref{eq21}}), we have that $F(0) = \infty$ and $F(\infty)=-\infty$. Therefore, there exists a unique $\lambda^*$ with $F(\lambda^*) =
0$.

Furthermore, as shown in \cite{frac_program}, in a concave fractional program, any local maximum is a global maximum. Therefore, the derived
$\lambda^*$ is global optimal.

\bibliographystyle{IEEEtran}
\bibliography{reference}

\end{document}